\documentclass[amsmath,amssymb,article,preprints,article,accept,moreauthors,pdftex,10pt,a4paper]{revtex4-1}

\usepackage[utf8x]{inputenc}
\usepackage{ucs}

\usepackage{booktabs}

\usepackage{verbatim}
\usepackage{amsfonts}
\usepackage{dcolumn}% Align table columns on decimal point
\usepackage{bm}% bold math
\usepackage[usenames]{color}
\usepackage{multirow}
\usepackage{graphicx}
\usepackage{hyperref}
\usepackage{subfig}
\usepackage{booktabs}
\usepackage{xcolor}
\usepackage[normalem]{ulem}

\usepackage{ucs}
\usepackage{amssymb}
\usepackage{makeidx}
\usepackage{diagbox}
\usepackage{tabularx}
\usepackage{amsthm}

\usepackage{float} % Allows putting an [H] in \begin{figure} to specify the exact location of the figure
\usepackage{wrapfig} % Allows in-line images such as the example fish picture

\usepackage{upgreek} % para poner letras griegas sin cursiva.
\usepackage{cancel} % para tachar.
\usepackage{mathdots} % para el comando \iddots
\usepackage{mathrsfs} % para formato de letra
\graphicspath{{figures/}} % Specifies the directory where pictures are stored

\begin{document}	
%\preprint{AIP/123-QED}
\title{Analytical Shannon information entropies for all discrete multidimensional hydrogenic states}
% Force line breaks with \\

\author{Irene V. Toranzo}
%\homepage{http://www.Second.institution.edu/~Charlie.Author.}
\affiliation{%
	Departamento de Matem\'atica Aplicada, Universidad Rey Juan Carlos, 28933 Madrid, Spain%\\This line break forced% with \\
}%

\author{David Puertas-Centeno}
%\homepage{http://www.Second.institution.edu/~Charlie.Author.}
\affiliation{%
	Departamento de Matem\'atica Aplicada, Universidad Rey Juan Carlos, 28933 Madrid, Spain%\\This line break forced% with \\
}%

\author{Nahual Sobrino}
%\homepage{http://www.Second.institution.edu/~Charlie.Author.}
\affiliation{%
	Donostia International Physics Center, 20018 Donostia, Gipuzkoa and
	Nano-Bio Spectroscopy Group and European Theoretical Spectroscopy Facility (ETSF), Dpto. de Fisica de Materiales, Universidad del Pa\'{\i}­s Vasco UPV/EHU, 20018 San Sebasti\'{a}n, Spain%\\This line break forced% with \\
}%

\author{Jes\'us S. Dehesa}
\email[]{dehesa@ugr.es}
%\homepage{http://www.Second.institution.edu/~Charlie.Author.}
\affiliation{%
	Departamento de F\'{\i}sica At\'{o}mica, Molecular y Nuclear, Universidad de Granada, 18071 Granada, Spain and
	Instituto Carlos I de F\'{\i}sica Te\'orica y Computacional, Universidad de Granada, 18071 Granada, Spain%\\This line break forced% with \\
}%

\date{\today}% It is always \today, today,

\begin{abstract}
The entropic uncertainty measures of the multidimensional hydrogenic states quantify the multiple facets of the spatial delocalization of the electronic probability density of the system. The Shannon entropy is the most adequate uncertainty measure to quantify the electronic spreading and to mathematically formalize the Heisenberg uncertainty principle, partially because it does not depend on any specific point of their multidimensional domain of definition.
In this work the radial and angular parts of the Shannon entropies for all the discrete stationary states of the multidimensional hydrogenic systems are obtained from first principles; that is, they are given in terms of the states' principal and magnetic hyperquantum numbers $(n,\mu_1,\mu_2,\ldots,\mu_{D-1})$, the system's dimensionality $D$ and the nuclear charge $Z$ in an analytical, compact form. Explicit expressions for the total Shannon entropies are given for the quasi-spherical states, which conform a relevant class of specific states of the $D$-dimensional hydrogenic system characterized by the hyperquantum numbers $\mu_1=\mu_2\ldots=\mu_{D-1}=n-1$, including the ground state.
\end{abstract}

\keywords{Entropic uncertainty; Shannon entropy of multidimensional hydrogenic states; dimensional dependence of Shannon entropy.}
%\submitto{\pdf}

\maketitle

\section{ INTRODUCTION}
The entropic uncertainty measures have been shown to be much more adequate than the Heisenberg ones (i.e., those based on the standard deviation and its moment-type generalizations \cite{zozor2011}) not only to formulate the fundamental uncertainty principle of quantum mechanics \cite{bialynicki1975,bialynicki2006,zozor2008,rudnicki2012} and for the development of quantum information and computation \cite{adesso2018,nielsen2010,sen2011,angulo2011,dehesa2011,rudnicki2018}, but also to describe the physical and chemical properties of multidimensional quantum systems. The latter is specially so in quantum chemistry where $D$-dimensional notions and techniques play a very relevant role as the chemists Dudley R. Herschbach (Nobel Prize 1986) \cite{Herschbach1993} and Gerhard Ertl (Nobel Prize 2007) and their collaborators have shown. For example, in chemistry two dimensions are often better than three, since surface-bound reactions can be proved in greater detail than those in a liquid solution, as Ertl et al illustrated in their pioneering contributions to the field of surface chemistry \cite{Cox1985,Jakubith1985,Beta1985}.
 Most efforts have been focussed on the entropic information entropies of R\'enyi and Shannon types and its variations \cite{renyi1961,aczel1975,leonenko2008,shannon1949,shannon1993}, basically because they have underpinned fundamental progress in a great diversity of scientific and technological fields such as applied mathematics, theoretical physics, quaand information technologies, including electronic correlations, complexity theory, statistical mechanics, quantum information theory, telecomunications, biochemical phenomena and quantum information processing \cite{dehesa2001,gadre2002,sen2011,adesso2018,esquivel2015b,esquivel2015c,toranzo2015b,dulieu2018}.\\

%\cite{dehesa2001,gadre2002,sen2011,adesso2018,gonzalez2003,esquivel2010,lopez2010,esquivel2012,esquivel2015b,esquivel2015c,toranzo2015b,esquivel2016a,esquivel2016b,esquivel2016c,dulieu2018}.

 The analytical determination of these entropic quantities is a serious task \cite{dehesa2011}, basically because the quantum probability density of these systems is not generally known, except for some extreme cases and/or some special states (high-dimensional and high-energy states) of reference systems of harmonic and hydrogenic types \cite{gallup1959,nieto1979,yanez1994,aquilanti1997,coletti2013}. The latter is because the probability density of these reference systems can be expressed by means of special functions of mathematical physics \cite{nikiforov1988,olver2010}, so that the aforementioned entropies are given by integral functionals of hypergeometric orthogonal polynomials \cite{yanez1994}. Even so, these entropic functionals have been calculated only for the ground and the first few lowest-lying states \cite{gadre1985,bhattacharya1998,ghosh2000,gadre2002} and for the extreme high-energy (i.e., Rydberg) \cite{aptekarev2016,dehesa2017,toranzo2016b,toranzo2016c} and  high-dimensional (i.e., pseudoclassical) \cite{puertas2017b,puertas2017,sobrino2017,dehesa2019} states of harmonic and Coulomb systems by use of fine asymptotics of orthogonal polynomials \cite{aptekarev1996,temme2017,aptekarev2010bis}. 

Recently, the R\'enyi uncertainty measures of the $D$-dimensional harmonic \cite{puertas2018bis} and hydrogenic \cite{puertas2018} systems have been analytically determined for all ground and excited quantum states from first principles; that is, directly in terms of $D$, the potential strength and the states' hyperquantum numbers. A similar achievement has been obtained this year for the Shannon entropy of the multidimensional harmonic states \cite{toranzo2019}. This has been possible basically because the harmonic wavefunctions can be factorized in Cartesian coordinates. Such phenomenon does not occur for hydrogenic states, so that a procedure similar to the harmonic case cannot be used to determine the explicit expression for the Shannon entropy of the multidimensional hydrogenic system; and, moreover, no alternative approach has been proposed up until now. This is amazing because, for example, the three-dimensional hydrogen is the simplest and most abundant element in nature; this element, formed right after the Big Bang, accounts for nine out of ten atoms in the universe, and nearly three-quarters of its mass; and, among many other important properties, it was the first atom trapped in the quest to form a Bose-Einstein condensate (BEC) and, moreover, the densest BEC is that of hydrogen to date \cite{dulieu2018}. \\

In this work we tackle the analytical determination of the uncertainty measures of  Shannon type for the $D$-dimensional hydrogenic systems in terms of the space dimensionality, the Coulomb strength (i.e., the nuclear charge $Z$) and the $D$ principal and magnetic hyperquantum numbers $(n,\mu_1,\mu_2,\ldots,\mu_{D-1})$ which characterize their quantum-mechanically allowed states. This is a far more difficult problem than the Heisenberg-like \cite{ray1988,drake1990,andrae1997,tarasov2004,dehesa2010,aptekarev2010b,zozor2011,toranzo2016} and Fisher information \cite{romera2006} cases, not only analytically but also numerically. The latter is basically because a naive numerical evaluation using quadratures is not convenient due to the increasing number of integrable singularities when the principal hyperquantum number is increasing, which spoils any attempt to achieve reasonable accuracy even for rather small hyperquantum number \cite{buyarov2004}. Up until now, the Shannon entropy of the multidimensional hydrogenic systems has been only calculated in a compact form at the high-dimensional (pseudoclassical) \cite{puertas2017b,dehesa2019} and high-energy (Rydberg) \cite{lopez2013,toranzo2016b,toranzo2016c} limits by use of modern asymptotical techniques of the Laguerre and Gegenbauer  polynomials which control the state's wavefunctions in position and momentum spaces \cite{aptekarev2016,temme2015,temme2017}.  It remains the exact calculation of this position-space entropic uncertainty measure for all discrete stationary  $D$-dimensional hydrogenic states. This is the aim of the present work.\\

The structure of the manuscript is the following. In Sec. \ref{sec:hydrodesc} the Shannon entropy for a $D$-dimensional probability density is defined, and then the wavefunctions of the hydrogenic states in the $D$-dimensional configuration space are briefly described so as to express the associated probability densities. Then, since the associated probability density is factorized in two radial and angular parts, one realizes that the total hydrogenic Shannon entropy can be expressed as the sum of two parts, the radial and the angular Shannon entropies. These two quantities are given in the form of logarithmic functionals of Laguerre orthogonal polynomials and hyperspherical harmonics, respectively. In Sec. \ref{sec:radsha} the radial hydrogenic Shannon entropy  is analytically determined by its decomposition into four integral functionals which are calculated in a compact form in terms of a particular instance of the multivariate Lauricella function of type A of $s$ variables and $2s+1$ parameters $F_{A}^{(s)}(x_{1},\ldots,x_{s})$ evaluated at unity; see e.g. \cite{olver2010,srivastava1985}. In Sec. \ref{sec:angsha} the angular hydrogenic Shannon entropy is expressed in a compact form by means of the $r$-variate Srivastava--Daoust function \cite{srivastava1988,sanchez2013}, $F_{1:1;\ldots;1}^{1:2;\ldots;2}(x_1,\ldots,x_r)$, evaluated at unity. In Section \ref{sec:totsha} the total hydrogenic Shannon entropy is discussed and some concluding remarks and open problems are given, respectively. Applications to the relevant quasi-spherical $D$-dimensional hydrogenic states are done; this specific class of states, which include the ground state, are characterized by the angular hyperquantum numbers $\mu_1=\mu_2\ldots=\mu_{D-1}=n-1$. Finally, in Sec. \ref{sec:conclu} conclusions are given and some open problems are pointed out.
\\
 
%%%%%%%%%%%%%%%%%%%%%%%%%%%%%%%%%%%%%%%%%%
\section{ MULTIDIMENSIONAL HYDROGENIC STATES: A SHANNON ENTROPIC VIEW}
\label{sec:hydrodesc}
In this section we first give the quantum-mechanical probability density $\rho(\vec{r})$ of the stationary states of the Coulomb potential $V_{D}(r) = - \frac{Z}{r},$ which characterizes a $D$-dimensional hydrogenic system. The  position vector $\vec{r}  =  (x_1 ,  \ldots  , x_D)$ in hyperspherical units  is  given as $(r,\theta_1,\theta_2,\ldots,\theta_{D-1})      \equiv
(r,\Omega_{D-1})$, $\Omega_{D-1}\in S^{D-1}$, so that the radial variable $r \equiv |\vec{r}| = \sqrt{\sum_{i=1}^D x_i^2}
\in [0 , \: +\infty)$  and $x_i =  r \left(\prod_{k=1}^{i-1}  \sin \theta_k
\right) \cos \theta_i$ for $1 \le i \le D$
%\sin\theta_1 \ldots \sin\theta_{k-1}  \cos\theta_k$,
and with $\theta_i \in [0 , \pi), i < D-1$, $\theta_{D-1} \equiv \phi \in [0 ,  2
\pi)$. Atomic units (i.e., $\hbar = m_e = e = 1$) are used throughout the paper. Then, we analyze the total electronic spreading of the system by means of the Shannon entropy of $\rho(\vec{r})$, which is defined \cite{shannon1949,shannon1993} as 
\begin{equation}
\label{shaen1}
S[\rho,D]:=-\int_{\Delta} \rho(\vec{r})\log\rho(\vec{r})\, d\,\vec{r},\quad \Delta \subseteq \mathbb{R^{D}}.
\end{equation}
It is well known that this quantity is a limiting case \cite{aczel1975} of the R\'enyi entropy, $R_{q}[\rho] = \frac{1}{1-q} \log \int_{\mathbb{R}^{D}}  [\rho(\vec{r})]^q \,d\vec{r}$, $q > 1$, so that $S[\rho] = \lim_{q\rightarrow 1} R_{q}[\rho]$. We start by solving the associated Schr\"{o}dinger equation of a $D$-dimensional ($D \geqslant 2$) spinless particle moving in the spherically-symmetric potential $V_{D}(r)$, as already described by various authors (see e.g. \cite{nieto1979,yanez1994,aquilanti1997,dehesa2010,coletti2013}), obtaining the eigenvalues $E=\lambda\left(2n+l+\frac D2\right)$, and the eigenfunctions 
\begin{eqnarray}
\label{wavpos}
\Psi_{n,l,\{\mu\}}(\vec{r}) &=&
N_{n,l}\left(\frac{r}{\lambda}\right)^{l}e^{-\frac{r}{\lambda}}L_{n-l-1}^{(2l+D-2)}\left(\frac{r}{\lambda}\right) \times \mathcal{Y}_{l,\{\mu\}}(\Omega_{D-1})\nonumber\\
&=& N_{n,l}\left[\frac{\omega_{2L+1}(\tilde{r})}{\tilde{r}^{D-2}}\right]^{1/2}L_{\eta-L-1}^{(2L+1)}(\tilde{r}) \times \mathcal{Y}_{l,\{\mu\}}(\Omega_{D-1}) \nonumber\\
&\equiv&\mathcal{R}_{n,l}(r) \times {\cal{Y}}_{l,\{\mu\}}(\Omega_{D-1}),
\end{eqnarray}
with 
\begin{eqnarray}
\eta &=& n + \frac{D-3}{2}, \quad n=1,2,3,\ldots\nonumber \\
\label{eq:1}
L &=& l+\frac{D-3}{2}, \quad l = 0,1,2,\ldots\nonumber \\
\label{eq:2}
\tilde{r} &=& \frac{r}{\lambda}\quad \text{with}\quad \lambda =\frac{\eta}{2Z},
\label{3}
\end{eqnarray}
 where the integer parameters
$(n,l,\{\mu\})\equiv(n,l\equiv\mu_1,\mu_2,\ldots,\mu_{D-1})$ denote the hyperquantum numbers associated to the variables $(r,\theta_1,\theta_2,\ldots,\theta_{D-1})$, which may take all values consistent with the inequalities $l\equiv\mu_1\geq\mu_2\geq\ldots\geq \left|\mu_{D-1} \right| \equiv \left|m\right|\geq 0$. The symbol  $\omega_{\alpha}(x) =x^{\alpha}e^{-x}, \, \alpha=2l+D-2$
  is the weight function of the orthogonal and orthonormal Laguerre polynomials of degree $n$ and parameter $\alpha$, here denoted by $L_{n}^{(\alpha)}(x)$ and $\widehat{L}_{n}^{(\alpha)}(x)$, respectively and
\begin{equation}
N_{n,l} = \lambda^{-\frac{D}{2}}\left\{\frac{(\eta-L-1)!}{2\eta(\eta+L)!}\right\}^{\frac{1}{2}}=\left\{\left(\frac{2Z}{n+\frac{D-3}{2}}\right)^{D}\frac{(n-l-1)!}{2\left(n+\frac{D-3}{2}\right)(n+l+D-3)!}  \right\}^{\frac{1}{2}}
\label{4}
\end{equation}
represents the normalization constant which ensures the unit norm of the wavefunction.
Note that the wavefunctions of the quantum-mechanically allowed stationary hydrogenic state are characterized by the $D$ hyperquantum numbers $(n,l,\{\mu\})$, the space dimensionality $D$ and the Coulomb strength $Z$. Moreover, they have a radial part $\mathcal{R}_{n,l}(r)$ and an angular part ${\cal{Y}}_{l,\{\mu\}}(\Omega_{D-1})$. The angular eigenfunctions are the hyperspherical harmonics, $\mathcal{Y}_{l,\{\mu\}}(\Omega_{D})$, defined as
\begin{eqnarray}
\label{eq:hyperspherarm}
\mathcal{Y}_{l,\{\mu\}}(\Omega_{D-1}) &=& \mathcal{N}_{l,\{\mu\}}e^{im\phi} \times \prod_{j=1}^{D-2}C^{(\alpha_{j}+\mu_{j+1})}_{\mu_{j}-\mu_{j+1}}(\cos\theta_{j})(\sin\theta_{j})^{\mu_{j+1}}
\end{eqnarray}
with $\alpha_{j}= (D-j-1)/2$ and the normalization constant
\begin{equation}
\label{eq:normhypersphar}
\mathcal{N}_{l,\{\mu\}}^{2} = \frac{1}{2\pi}\times\nonumber\\
\prod_{j=1}^{D-2} \frac{(\alpha_{j}+\mu_{j})(\mu_{j}-\mu_{j+1})![\Gamma(\alpha_{j}+\mu_{j+1})]^{2}}{\pi \, 2^{1-2\alpha_{j}-2\mu_{j+1}}\Gamma(2\alpha_{j}+\mu_{j}+\mu_{j+1})},\nonumber\\
\end{equation}
where the symbol $C^{(\lambda)}_{n}(t)$ denotes the Gegenbauer polynomial of degree $n$ and parameter $\lambda$. Moreover, these hyperfunctions satisfy the orthonomalization condition
\begin{equation}\label{eq:normalizacionY}
\int_{\mathcal{S}_{D-1}} d\Omega_{D-1 } {\cal{Y}}^{*}_{l', \left\lbrace \mu' \right\rbrace}(\Omega_{D-1})
{\cal{Y}}_{l, \left\lbrace \mu \right\rbrace}\left( \Omega_{D-1} \right) = \delta_{l,l'} \delta_{\left\lbrace \mu \right\rbrace,
\left\lbrace \mu' \right\rbrace}
\end{equation}
Then, the position probability density of a $D$-dimensional hydrogenic state with hyperquantum numbers, $(n,l,\{\mu\})$, is given by the squared modulus of the position eigenfunction $\Psi_{\eta,L,\{\mu\}}(\vec{r})$ as follows 
\begin{eqnarray}
\label{denspos}
\rho_{n,l,\{\mu\}}(\vec{r}) &=& \mathcal{R}_{n,l}^2(r)\times |\mathcal{Y}_{l,\{\mu\}}(\Omega_{D-1})|^{2}  \nonumber \\
 &=& N_{\eta,l}^{2}\left[\frac{\omega_{2L+1}(\tilde{r})}{\tilde{r}^{D-2}}\right][L_{\eta-L-1}^{(2L+1)}(\tilde{r})]^{2} \times |\mathcal{Y}_{l,\{\mu\}}(\Omega_{D-1})|^{2} \nonumber \\
&=& N_{n,l}^{2}\tilde{r}^{2l}e^{-\tilde{r}}[L_{n-l-1}^{(2l+D-2)}(\tilde{r})]^{2} \times |\mathcal{Y}_{l,\{\mu\}}(\Omega_{D-1})|^{2}\nonumber \\
&\equiv& \rho_{n,l}(\tilde{r})\times |\mathcal{Y}_{l,\{\mu\}}(\Omega_{D-1})|^{2}.
\end{eqnarray}

Note that the wavefunctions are duly normalized so that  $\int \rho_{\eta,l, \left\lbrace \mu \right\rbrace }(\vec{r}) d\vec{r} =1$, where the $D$-dimensional volume element $d\vec{r} = r^{D-1} dr\,d\Omega_{D-1}$ with
$$
d\Omega_{D-1}=\left(\prod_{j=1}^{D-2} (\sin \theta_j)^{2 \alpha_j} d\theta_j\right) d\theta_{D-1},
$$
and the normalization to unity of the hyperspherical harmonics given by $\int |\mathcal{Y}_{l,\{\mu \}}(\Omega_{D})|^{2}d\Omega_{D} = 1$ was taken into account.

The total spreading of this density is known to be best quantified by its Shannon entropy defined by Eq. \eqref{shaen1}, which can be decomposed as 
\begin{equation}
\label{shaen2}
S[\rho_{n,l,\{\mu \}},D] = S[\mathcal{R}_{n,l},D]  + S[\mathcal{Y}_{l,\{\mu \}},D] ,
\end{equation}
where
\begin{equation}
\label{shaenr1}
S[\mathcal{R}_{n,l},D]=  - \int_{0}^{\infty} r^{D-1}\,\rho_{nl}(r)\log \rho_{nl}(r)\, dr 
\end{equation}
and
\begin{equation}
\label{shaena2}
S[\mathcal{Y}_{l,\{\mu \}},D]  = -\int_{\mathcal{S}_{D-1}} |\mathcal{Y}_{l,\{\mu \}}(\Omega_{D-1})|^{2}\log |\mathcal{Y}_{l,\{\mu \}}(\Omega_{D-1})|^{2}\, d\Omega_{D-1} 
\end{equation}
denote the radial and angular parts of the Shannon entropy, respectively, which will be calculated in the next two sections for any stationary state $(n,l,\{\mu\})$. Note that the radial part depends on the hyperquantum numbers $\{n,l\equiv\mu_{1}\}$ only, and the angular part does not depend on the principal hyperquantum number $n$, but only on the magnetic hyperquantum numbers $\{\mu_{i},i=1,\dots,D-1\}$.

%%%%%%%%%%%%%%%%%%%%%%%%%%%%%%%%%%%%%%%%%%
\section{ THE RADIAL SHANNON ENTROPY}
\label{sec:radsha}
The radial part of the Shannon entropy of the hydrogenic density, according Eq. \eqref{denspos} and Eq. \eqref{shaenr1}, is given by
\begin{equation}
\label{serad}
S[\mathcal{R}_{n,l},D]=-\int_{0}^\infty r^{D-1}\mathcal{R}_{nl}^{2}(r)\,\log\,\left(\mathcal{R}_{nl}^{2}(r)\right)dr
\end{equation}
This quantity is known \cite{dehesa2010,lopez2013,toranzo2016b} to have the expression
\begin{equation}\label{eqI_cap1:SroAsin}
S[\mathcal{R}_{n,l},D] = 2D \log n +(2-D) \log 2+ \log \pi+D-3-D \log Z + o(1); \quad n \to \infty,
\end{equation}
for the high-energy (Rydberg) $D$-dimensional hydrogenic state characterized by the hyperquantum numbers $(n,l,\{\mu\})$ with a large principal quantum number $n$. Moreover, it has been conjectured and numerically shown to be correct \cite{puertas2017b} that
\begin{equation} \label{conject}
   S[\mathcal{R}_{n,l},D]\sim 2D \log\, D-D \log \left(4Z\right), \quad D \to \infty,
   \end{equation}
which gives the radial Shannon entropy for the high-dimensional hydrogenic states. The symbol $A \sim B$ means that $A/B \to 1$.\\
In this section we will extend these results by calculating the radial Shannon entropy $S[\mathcal{R}_{n,l},D]$ for all discrete stationary $D$-dimensional hydrogenic state $(n,l,\{\mu\})$. Then, according to Eq. \eqref{wavpos}, we have from Eq. \eqref{serad} that the radial Shannon entropy for an arbitrary stationary state characterized by the hyperquantum numbers $(n,l,\{\mu\})$ can be expressed as
\begin{align}
\label{shaenr2}
S[\mathcal{R}_{n,l},D] &=  - N_{n,l}^{2}\lambda^{D}\int_{0}^{\infty} x^{2l+D-1}e^{-x}[L_{n-l-1}^{(2l+D-2)}(x)]^{2} \log \left\{N_{n,l}^{2}x^{2l}e^{-x}[L_{n-l-1}^{(2l+D-2)}(x)]^{2}\right\}\, dx  \nonumber \\
& = - N_{n,l}^{2}\lambda^{D} \Bigg\{\log(N_{n,l}^{2}) \int_{0}^{\infty} x^{2l+D-1}e^{-x}[L_{n-l-1}^{(2l+D-2)}(x)]^{2} \, dx  \nonumber \\
& + 2l \int_{0}^{\infty} x^{2l+D-1}e^{-x}[L_{n-l-1}^{(2l+D-2)}(x)]^{2} \log x\, dx  - \int_{0}^{\infty} x^{2l+D}e^{-x}[L_{n-l-1}^{(2l+D-2)}(x)]^{2} \, dx  \nonumber \\
& + \int_{0}^{\infty} x^{2l+D-1}e^{-x}[L_{n-l-1}^{(2l+D-2)}(x)]^{2} \log[L_{n-l-1}^{(2l+D-2)}(x)]^{2}\, dx \Bigg\}\nonumber \\
& \equiv  - N_{n,l}^{2}\lambda^{D} \left( \log(N_{n,l}^{2})\,\mathcal{I}_{1} + 2l\,\mathcal{I}_{2}-\mathcal{I}_{3} + \mathcal{I}_{4}   \right).
\end{align}
To determine the integrals $\mathcal{I}_{i}, i = 1-4,$ we will use the following accessory integral of the Laguerre polynomials (which can be obtained by linearizing the powers of these polynomials following the procedure described in \cite{puertas2018})
\begin{align}
\label{accint1}
\mathcal J &= \int_{0}^{\infty} x^{\beta} e^{-x} [L_{m}^{(\alpha)}(x)]^{2q} \, dx = \Gamma(\beta+1)\binom{m+\alpha}{m}^{2q} F_{A}^{(2q)} \left(\begin{array}{cc}
		\beta+1 ; -m, \ldots, -m& \\
		&; 1, \ldots, 1\\
		\alpha+1, \ldots, \alpha+1 & \\
	\end{array}\right),
\end{align}
where $F_{A}^{(s)}(x_{1},\ldots,x_{r})$ is the multivariate Lauricella function of type A of $s$ variables and $2s+1$ parameters; see e.g. \cite{olver2010,srivastava1985}. The general hypergeometric function $F_{A}^{(s)}(x_{1},\ldots,x_{r})$ is given by

 \begin{equation}
 \label{laurifunc}
 %\resizebox{\linewidth}{!} 
 % {
%      $
 F_{A}^{(s)}\left( \begin{array}{cc}
  							a; b_{1}, \ldots, b_{s} & \\
  																			&; x_{1},\ldots, x_{s}\\
  							c_{1}, \ldots, 	c_{s} & \\
  							\end{array}\right) =\sum_{j_{1},\ldots, j_{s}=0}^{\infty} \frac{(a)_{j_{1}+\ldots+j_{s}}(b_{1})_{j_{1}} \cdots (b_{s})_{j_{s}}}{(c_{1})_{j_{1}} \cdots (c_{s})_{j_{s}}} \frac{x_{1}^{j_{1}}\cdots x_{s}^{j_{s}} }{j_{1}!\cdots j_{s}! }.
  %$
  %}
 \end{equation}
where the Pochhammer symbol $(z)_a = \frac{\Gamma(z+a)}{\Gamma(z)}$.

To solve the integrals $\mathcal{I}_{1}$ and $\mathcal{I}_{3}$ we use Eq. \eqref{accint1} with the sets of parameters $(\beta=2l+D-1, \alpha=2l+D-2,m=n-l-1)$ and  $(\beta=2l+D, \alpha=2l+D-2,m=n-l-1)$, obtaining the expressions
	\begin{align} 
	\label{I1}
		\mathcal I_{1} &= \Gamma\left(2l+D\right)\,\binom{n+l+D-3}{n-l-1}^2\, F_{A}^{(2)}\left( \begin{array}{cc}
			2l+D; -n+l+1,-n+l+1  & \\
			&; 1,1\\
			2l+D-1,\quad 2l+D-1  & \\
		\end{array}\right)\nonumber \\
		&=\Gamma\left(2l+D\right)\,\binom{n+l+D-3}{n-l-1}^2\,\sum_{i,j=0}^{n-l-1}\frac{(2l+D)_{i+j}\,(-n+l+1)_i\,(-n+l+1)_j}{(2l+D-1)_i\,(2l+D-1)_j\,i!\,j!}\nonumber\\
		  &=(n-l)_{2l+D-2}^2 \,\sum_{i,j=0}^{n-l-1}\,c_{i,j}(n,l,D),
	  	\end{align} 
and
	\begin{align}
	\label{I3}
		\mathcal I_{3}  &= \Gamma\left(2l+D+1\right)\,\binom{n+l+D-3}{n-l-1}^2\, F_{A}^{(2)}\left( \begin{array}{cc}
			2l+D+1; -n+l+1,-n+l+1  & \\
			&; 1,1\\
			2l+D-1,\quad 2l+D-1  & \\
		\end{array}\right)\nonumber \\&= \Gamma\left(2l+D+1\right)\,\binom{n+l+D-3}{n-l-1}^2\,\sum_{i,j=0}^{n-l-1}\frac{(2l+D+1)_{i+j}\,(-n+l+1)_i\,(-n+l+1)_j}{(2l+D-1)_i\,(2l+D-1)_j\,i!\,j!}\nonumber \\&= (n-l)_{2l+D-2}^2 \,\sum_{i,j=0}^{n-l-1}\,c_{i,j}(n,l,D)\,(2l+D+i+j)\quad
	\end{align}
respectively, and where the coefficients $c_{i,j}(n,l,D)$ are given by
	\begin{equation}
	\label{coefficients}
	c_{i,j}(n,l,D)=\frac{(2l+D-1+i)_{1+j}}{\Gamma(2l+D-1+j)}\frac{(-n+l+1)_i(-n+l+1)_j}{i!j!} 
	\end{equation} 
To calculate the integral $\mathcal{I}_{2}$ we realize that it is related to the integral $\mathcal J$ defined in Eq. \eqref{accint1} as follows
\begin{align}
\label{i2sol1}
\mathcal I_2=\left.\frac{d}{d\beta}\mathcal J \right|_{(\beta=2l+D-1,\,\alpha=2l+D-2,\,m=n-l-1,\,q=1)} &= \int_{0}^{\infty} x^{2l+D-1} \log(x) e^{-x} \left[L_{n-l-1}^{(2l+D-2)}(x)\right]^2\, dx \nonumber \\
\end{align}
Thus, by calculating the derivative in Eq. \eqref{accint1}, and substituting in Eq. \eqref{i2sol1}, we find the following expression for $\mathcal{I}_{2}$,
\begin{align}
\label{I2}
\mathcal I_2&=\Gamma(2l+D)\,\binom{n+l+D-3}{n-l-1}^2\,\sum_{i,j=0}^{n-l-1}\frac{(2l+D)_{i+j}\,(-n+l+1)_i\,(-n+l+1)_j}{(2l+D-1)_i\,(2l+D-1)_j\,i!\,j!}\,\psi(2l+D+i+j),\nonumber
\\ &= (n-l)_{2l+D-2}^2 \,\sum_{i,j=0}^{n-l-1}\,c_{i,j}(n,l,D)\,\psi(2l+D+i+j)
\end{align} 
where the digamma function $\psi(x)= \frac{\Gamma^{'}(x)}{\Gamma(x)}$. To compute $\mathcal{I}_{4}$ we use that
 \begin{align}
 \label{i4sol1}
 \frac{d}{dq}\mathcal J\Bigg|_{\beta=\alpha+1,\, q=1} &= \int_{0}^{\infty} x^{\alpha+1} e^{-x} [L_{m}^{(\alpha)}(x)]^{2} \log [L_{m}^{(\alpha)}(x)]^{2}\, dx = \frac{d}{dq} \Bigg[\Gamma(\alpha+2)\binom{m+\alpha}{m}^{2q}\nonumber \\
 & \hspace{4cm}\times F_{A}^{(2q)} \left(\begin{array}{cc}
 \alpha +2 ; -m, \ldots, -m& \\
 &; 1, \ldots, 1\\
 \alpha+1, \ldots, \alpha+1 & \\
 \end{array}\right)\Bigg]\Bigg|_{q=1},
 \end{align}
 with $\alpha = 2l+D-2$ and $m=n-l-1$ and where $F_{A}^{(r)}$ denotes the Lauricella function of $r$ variables and $2r+1$ parameters evaluated at unity as defined by Eq. \eqref{laurifunc}. Then, keeping in mind the Laguerre weight function $\omega_\alpha(x) = x^{\alpha} e^{-x}$, we have 
%\textcolor{red}{ \begin{align}
% \mathcal{I}_{4} &= \int_{0}^{\infty} x^{\alpha+1} e^{-x} [L_{m}^{(\alpha)}(x)]^{2} \log [L_{m}^{(\alpha)}(x)]^{2}\, dx  \\
% &= \int_{0}^{\infty} x \,\omega_\alpha(x)  [L_{m}^{(\alpha)}(x)]^{2} \log [L_{m}^{(\alpha)}(x)]^{2}\, dx  \nonumber\\
% & = \Gamma(\alpha+2)\binom{m+\alpha}{m}^{2}\left(\log\left[\binom{m+\alpha}{m}^{2}\right] F_{A}^{(2)} \left(\begin{array}{cc}
% \alpha +2 ; -m, -m& \\
% &; 1, 1\\
% \alpha+1, \alpha+1 & \\
% \end{array}\right)\right.\nonumber\\
% &+\left.\frac{d}{dq}\left[F_{A}^{(2q)} \left(\begin{array}{cc}
% \alpha +2 ; -m,\ldots, -m& \\
% &; 1,\ldots, 1\\
% \alpha+1,\ldots, \alpha+1 & \\
% \end{array}\right)\right]_{q=1}\right),
% \label{i4sol2}
% \end{align}
% or finally,}
 \begin{align}
 \label{I42}
 \mathcal{I}_{4} &= \int_{0}^{\infty} x^{2l+D-1} e^{-x} [L_{n-l-1}^{(2l+D-2)}(x)]^{2} \log [L_{n-l-1}^{(2l+D-2)}(x)]^{2}\, dx \nonumber \\
 & = \Gamma(2l+D)\binom{n+l+D-3}{n-l-1}\left(\log\left[\binom{n+l+D-3}{n-l-1}^{2}\right] F_{A}^{(2)} \left(\begin{array}{cc}
 2l+D ; -n+l+1, -n+l+1& \\
 &; 1, 1\\
 2l+D-1, 2l+D-1 & \\
 \end{array}\right)\right.\nonumber\\
 &+\left.\frac{d}{dq}\left[F_{A}^{(2q)} \left(\begin{array}{cc}
 2l+D ; -n+l+1,\ldots, -n+l+1& \\
 &; 1,\ldots, 1\\
 2l+D-1,\ldots, 2l+D-1 & \\
 \end{array}\right)\right]_{q=1}\right),
 \end{align}
 where $F_{A}^{(2)}(1,1)$ denotes the Appell function of second kind evaluated at unity. It is interesting to remark that the last term has the following integral representation
 \begin{align}
 	\label{der_FA}
\left.\frac{d}{dq}F_{A}^{(2q)} \begin{pmatrix}
   \alpha+2;-m,...,-m &  \\
     & ;1,...,1 \\
    \alpha+1,...,\alpha+1 &  \end{pmatrix}\right\vert_{q=1}
  =\frac{1}{\Gamma(\alpha+2)}\int_{0}^{\infty}&e^{-t}t^{\alpha+1}\left[ \textsubscript{1}F_{1}(-m; \alpha +1;t)\right]^{2}\nonumber\\
  &\times\log\left[ \textsubscript{1}F_{1}(-m; \alpha +1;t)\right]^{2}dt,
\end{align}
where the symbol $_1F_1(a;b;t)$ denotes the confluent hypergeometric or Kummer function \cite{olver2010}.
In fact, if we denote 
\begin{equation} \label{integralKummer}
\widetilde{\mathcal I}=\int_0^\infty e^{-t}t^{2l+D-1}\left[ \textsubscript{1}F_{1}(-n+l+1; 2l+D-1;t)\right]^{2}\,\log\left[ \textsubscript{1}F_{1}(-n+l+1; 2l+D-1;t)\right]^{2}dt
\end{equation}
we can write
\begin{equation}
\label{I4}\mathcal I_4= \log\left[\binom{n+l+D-3}{n-l-1}^{2}\right]\,\mathcal I_1+\binom{n+l+D-3}{n-l-1}\,\widetilde{\mathcal I}.
\end{equation}
Finally, using the equations \eqref{I1},\eqref{I2},\eqref{I3}, and \eqref{I4} and after some algebraic manipulations we have
\begin{equation}\label{radentropy}
S[\mathcal{R}_{n,l},D]=\frac{-1}{2\,\eta}\left[(n-l)_{2l+D-2}\sum_{i,j=0}^{n-l-1}c_{i,j}(n,l,D)\,\mathcal K_{i,j}(Z,n,l,D)+\frac{\widetilde{\mathcal I}}{\Gamma(2l+D-1)}\right],
\end{equation}
with 
\begin{equation}\label{radaux}
 \mathcal K_{i,j}(Z,n,l,D)=D\log\left(\frac{2\,Z}{\eta}\right)+\log\left(\frac{(n-l)_{2l+D-2}}{2\eta\,\Gamma(2l+D-1)}\right)+2\,l\,\psi(2l+D+i+j)-(2l+D+i+j),
\end{equation}
and where $c_{i,j}(n,l,D)$ is defined in Eq. \eqref{coefficients}. From Eqs. \eqref{radentropy}-\eqref{radaux} it must be possible to obtain the expressions \eqref{eqI_cap1:SroAsin} and \eqref{conject} corresponding to the two extreme cases $n \to \infty$ and $D \to \infty$, respectively; however, there is a special mathematical problem, not yet solved, related to the corresponding limiting cases of the entropy-like integrals of the involved Kummer function $\widetilde{\mathcal I}$ defined by eq. \eqref{integralKummer}.\\

The expression \eqref{radentropy}, which gives the radial Shannon entropy $S[\mathcal{R}_{n,l},D]$ for all discrete stationary $D$-dimensional hydrogenic state $(n,l,\{\mu\})$, can be algebraically evaluated by means of symbolic programs such as Mapple and Mathematica and its variations (see e.g. \cite{koepf2014,gerdt2018} and references therein) in a straightforward manner. They can perform highly sophisticated algebraic tasks and, moreover, they are equipped for solving problems from mathematical analysis in a symbolic way, including the analytical evaluation of the generalized univariate and multivariate hypergeometric functions.\\

For some particular electronic states this physical entropy can be expressed in a simple and compact manner. For instance, the radial Shannon entropy for the hydrogenic $(n,l=n-1)$-states, is 

\begin{equation*}
S[\mathcal{R}_{n,n-1},D] = \log \left[ \left(\frac{\eta}{2}\right)^D\, \,e^{2\eta+1}\,\Gamma(2\eta+1) \right]-2(n-1) \,\psi (2\eta+1) -D\log \,Z
\end{equation*}
(with $\eta = n + \frac{D-3}{2}$) and, consequently, for the ground state $(n=1,0,\{0 \})$ we have

\begin{equation*}
	S[\mathcal{R}_{1,0},D] =\log\left((D-1)^D\,\Gamma(D)\right)-D\log \left(\frac{4}e\right)-D\log Z
\end{equation*}
which gives the known values $2-4\log2$ and $3-2\log 2$ in the two and three dimensional hydrogen ($Z=1$) cases, respectively.\\
Finally, for completeness, let us mention here that there exist other possible approaches to determine the radial Shannon entropy of the $D$-dimensional hydrogenic systems which are based on mathematical entropic notions of the involved Laguerre polynomials such as the entropy-like functionals and/or the logarithmic potentials of these polynomials \cite{dehesa2001,saff1997}. However, the analytical determination of these mathematical objects is not yet well explored and, in any case, it requires the use of the linearization coefficients of Laguerre polynomials and the knowledge of the zeros of Laguerre polynomials.

 %%%%%%%%%%%%%%%%%%%%%%%%%%%%%%%%%%%%%%%%%%
\section{ THE ANGULAR SHANNON ENTROPY}
 \label{sec:angsha}
 The angular Shannon entropy $S[\mathcal{Y}_{l,\{\mu \}},D]$ defined by Eq. \eqref{shaena2} denotes the entropy of the hyperspherical harmonics and quantifies the angular spatial uncertainty of the multidimensional single-particle systems. Up until now there is no closed expression for this quantity for arbitrary hyperquantum numbers $(l,\{\mu \})$ although a number of efforts have been done \cite{yanez1999,dehesa2007,dehesa2010,dehesa2015,dehesa2019}. In particular it has been conjectured \cite{puertas2017b} that this quantity satisfies
 \begin{equation} \label{limit}
 S[\mathcal{Y}_{l,\{\mu \}},D]\sim-\log\left(\Gamma\left(\frac D2\right)\right)+\frac D2\log{\pi}, \quad D \to \infty,
  \end{equation} 
for the high-dimensional case. Moreover, it has been shown \cite{yanez1999,dehesa2010} that the angular Shannon entropy $S[\mathcal{Y}_{l,\{\mu \}},D]$ can be expressed for $D \geq 2$ as
\begin{equation}\label{shang}
 S[\mathcal{Y}_{l,\{\mu \}},D] = B (l,\left\lbrace \mu \right\rbrace,D)+\sum_{j=1}^{D-2} E\left[ \tilde{{\cal{C}}}_{\mu_j-\mu_{j+1}}^{(\alpha_j+\mu_{j+1})}\right]
\end{equation}
in terms of the quantum numbers $(l,\{\mu\})$ and the dimensionality $D$ by means of the entropy of the orthonormal Gegenbauer polynomials, where
\begin{eqnarray}\label{eq:coefB}
B (l,\left\lbrace \mu \right\rbrace,D)&= \log\, (2\pi) -2 \sum^{D-2}_{j=1} \mu_{j+1}
\left[\psi(2\alpha_j+\mu_j+\mu_{j+1})\right. \nonumber \\
&\quad \left. -\psi(\alpha_j+\mu_j)- log\, 2 -\frac{1}{2 (\alpha_j+\mu_j)}\right],
\end{eqnarray}
and ${\tilde{\mathcal{C}}}_n^{(\lambda)}(x)$ denotes the Gegenbauer polynomial orthonormal with respect to the weight function
$\omega_\lambda(x)=(1-x^2)^{\lambda-\frac{1}{2}}$ on the interval $[-1,+1]$. The orthonormal Gegenbauer
polynomial ${\tilde{\mathcal{C}}}_n^{(\lambda)}(x)$ is related to the orthogonal Gegenbauer polynomial $C_n^{(\lambda)}(x)$
 by the relation
 \begin{equation}
{\tilde{\mathcal{C}}}_n^{(\lambda)}(x)=\frac{C_n^{(\lambda)}(x)}{h_n};\quad
\textrm{with}\quad h_n^2=\frac{2^{1-2\lambda} \pi \Gamma(n+2\lambda)}
{\left[\Gamma(\lambda) \right]^2(n+\lambda)n!}.
\end{equation}
Moreover, the symbol $E[y_n]$ denotes the Shannon-like entropy of the polynomials $y_n(x)$ orthogonal with respect to the weight function
$\omega(x)$ on the interval $(a,b)$ given by
\begin{equation}\label{integralE}
E[y_n]:= -\int_a^b \omega(x)\, y_n^2(x)\log \,y_n^2(x)\,  dx,
\end{equation}
Since the calculation of this entropy-like integral of Gegenbauer polynomials is a very difficult mathematical task (not yet accomplished, save for very particular cases), the explicit values of the angular entropy $S[\mathcal{Y}_{l,\{\mu \}},D]$ for any multi-index $(l,\left\lbrace \mu \right\rbrace)$ cannot be analytically carried out with Eqs. \eqref{shang}-\eqref{integralE} up until now.\\

In this section we show a different approach to calculate the angular Shannon entropy $S[\mathcal{Y}_{l,\{\mu \}},D]$ by using some recent results \cite{puertas2018} obtained for the exact R\'enyi entropy of the hydrogenic systems, $R_{q}[\rho_{n,l,\{\mu\}}] = \frac{1}{1-q} \log \int_{\mathbb{R}^{D}}  [\rho_{n,l,\{\mu\}}(\vec{r})]^q \,d\vec{r}$, $q > 1$, and the limiting relation between the Shannon and R\'enyi entropies mentioned above. Then, taking into account the relation between the angular Shannon entropy and the angular entropic moment of order q, $\Lambda_{l,\{\mu\}}(q)$, we can calculate the angular Shannon entropy of any non-relativistic  and spherically-symmetric multidimensional quantum system as
\begin{equation}
 	S[\mathcal{Y}_{l,\{\mu \}},D]= \lim_{q \to 1} \frac{1}{1-q} \log \Lambda_{q}[\mathcal{Y}_{ l,\{\mu\}},D,q] = -\left.\frac{d\,\Lambda_{q}[\mathcal{Y}_{ l,\{\mu\}},D,q]}{dq}\right\vert_{q=1},
 \end{equation}
where L'H\^{o}pital's rule has been used in the second equality. Then, with
 \begin{align}
 \label{angpart}
\Lambda_{q}[\mathcal{Y}_{ l,\{\mu\}},D,q] &= \int |\mathcal{Y}_{l,\{\mu\}}(\Omega_{D-1})|^{2q}\, d\Omega_{D-1}\nonumber\\
&= \left(2\pi^{\frac{D}{2}}\right)^{1-q}\frac{\Gamma\left(l+\frac{D}{2}\right)^{q}\Gamma\left(qm+1\right)}{\Gamma\left(ql+\frac{D}{2}\right)\Gamma\left(m+1\right)^{q}}\prod_{j=1}^{D-2}\mathcal{B}_{q}(D,\mu_{j},\mu_{j+1})\,\,\mathcal{G}_{q}(D,\mu_{j},\mu_{j+1}),
\end{align}
 where
 \begin{align}
 	\label{B_q}
\mathcal{B}_{q}(D,\mu_{j},\mu_{j+1})=\frac{1}{\left[\left(\mu_{j}-\mu_{j+1}\right)!\right]^{q}}\frac{(2\alpha_{j}+2\mu_{j+1}+1)_{2(\mu_{j}-\mu_{j+1})}^{q}}{(2\alpha_{j}+\mu_{j}+\mu_{j+1})_{\mu_{j}-\mu_{j+1}}^{q}}\frac{(q\mu_{j}+\alpha_{j}+1)_{q(\mu_{j}-\mu_{j+1})}}{(\mu_{j+1}+\alpha_{j}+1)_{\mu_{j}-\mu_{j+1}}^{q}}
\end{align}
and
\begin{align}
	\label{G_q}
\mathcal{G}_{q}(D,\mu_{j},\mu_{j+1})&=F_{1:1;\ldots;1}^{1:2;\ldots;2}\left(\begin{array}{cc}
 a_j:b_j,c_j; \ldots, b_j,c_j& \\
 &; 1, \ldots, 1\\
 d_j:e_j; \ldots, e_j & \\
 \end{array}\right)\nonumber\\
 &=\sum_{i_1,\ldots,i_{2q}}^{\mu_{j}-\mu_{j+1}}\frac{\left(a_j\right)_{i_1+\ldots i_{2q}}}{\left(d_j\right)_{i_1+\ldots i_{2q}}}\frac{\left(b_j\right)_{i_1}\left(c_j\right)_{i_1}\cdots\left(b_j\right)_{i_{2q}}\left(c_j\right)_{i_{2q}}}{\left(e_j\right)_{i_1}\cdots\left(c_j\right)_{i_{2q}}i_{1}!\cdots i_{2q}!}
\end{align}
 with $a_j = \alpha_j + q\mu_{j+1} + \frac{1}{2}$, $b_j =-\mu_j + \mu_{j+1}$, $c_j = 2\alpha_j + \mu_{j+1} + \mu_j$, $d_j = 2q\mu_{j+1} + 2\alpha_j + 1$ and $e_j = \alpha_j + \mu_{j+1}+\frac{1}{2}$. The symbol $F_{1:1;\ldots;1}^{1:2;\ldots;2}(x_1,\ldots,x_r)$ denotes the $r$-variate Srivastava--Daoust function \cite{srivastava1988,sanchez2013} defined as
\begin{eqnarray}
\label{daoust}
F_{1:1;\ldots;1}^{1:2;\ldots;2}\left( \begin{array}{cc}
a_0^{(1)}:\,a_1^{(1)},a_1^{(2)}; \ldots;a_r^{(1)},a_r^{(2)} & \\
&; x_1, \ldots, x_r\\
b_0^{(1)}:\,b_1^{(1)}; \ldots;b_r^{(1)}& \\
\end{array}\right) &=&\nonumber\\
=  \sum_{j_{1}, \ldots, j_r=0 }^{\infty} \frac{\left(a_0^{(1)}\right)_{j_1+\ldots+j_r}}{\left(b_0^{(1)}\right)_{j_1+\ldots+j_r}} \frac{\left(a_1^{(1)}\right)_{j_1}\left(a_1^{(2)}\right)_{j_1}\cdots\left(a_r^{(1)}\right)_{j_r}\left(a_r^{(2)}\right)_{j_r}}{\left(b_1^{(1)}\right)_{j_1}\cdots\left(b_r^{(1)}\right)_{j_r}}\frac{x_1^{j_1}x_2^{j_2}\cdots x_r^{j_r}}{j_1!j_2!\cdots j_r!}, & & \nonumber\\
\end{eqnarray}
Evaluating the limit $q \to 1$, the angular Shannon entropy becomes
\begin{align}\label{shang2}
\hspace{-1cm}S[\mathcal{Y}_{l,\{\mu\}},D]=\log {\left(2\pi^{\frac{D}{2}}\right)}+l\psi\left(l+\frac{D}{2}\right)&-m\psi\left(m+1\right)+\log{\left[\frac{\Gamma(m+1)}{\Gamma(l+\frac{D}{2})}\right]}\nonumber\\
&-\frac{d}{dq}\left.\left(\prod_{j=1}^{D-2}\mathcal{B}_{q}(D,\mu_{j},\mu_{j+1})\,\,\mathcal{G}_{q}(D,\mu_{j},\mu_{j+1})\right)\right\vert_{q=1}
\end{align}
%\textcolor{purple}{It would be desirable to have an integral expression for $\frac{d}{dq} \mathcal G_q$ similar to expression \eqref{der_FA} for the radial part remains as an open problem.}
%, due to  the function $\mathcal G_q$ given by the equation \eqref{G_q} is not derivable in $q$ for an arbitrary state, except for the case $\mu_j=\mu_{j+1}$ for what $\mathcal G_q\equiv\mathcal B_q\equiv 1,\quad \forall q>0$. }
The two previous approaches to calculate the angular Shannon entropy $S[\mathcal{Y}_{l,\{\mu \}},D]$, which are based on Eqs. \eqref{shang} and \eqref{shang2} respectively, are valid for arbitrary hyperquantum numbers $(l,\{\mu \})$. From them one can obtain much simpler expressions for particular electronic states. For instance, the angular Shannon entropy for the states $(l,\{l \})$ is 
\begin{equation}
S\left[ {\cal{Y}}_{l,\left\lbrace l \right\rbrace},D \right] = \log {\left(2\pi^{\frac{D}{2}}\right)}+l\,\left(\psi\left(l+\frac{D}{2}\right)-\psi\left(l+1\right)\right)+\log{\left[\frac{\Gamma(l+1)}{\Gamma(l+\frac{D}{2})}\right]}, \label{eqI_cap1:SYgrado0} 
\end{equation}
%((Se puede simplificar mÃ¡s usando los nÃºmeros armÃ³nicos))
%\begin{equation*}
%S\left[ {\cal{Y}}_{l,\left\lbrace l \right\rbrace},D \right] = \log {\left(2\pi^{\frac{D}{2}}\right)}+l\left(H_{l+\frac D2-1}-H_l\right)+\log{\left[\frac{\Gamma(l+1)}{\Gamma(l+\frac{D}{2})}\right]},\\ \label{eqI_cap1:SYgrado0} 
%\end{equation*}
where we have used that $\mathcal B_q(D,l,l)=\mathcal G_q(D,l,l)=1,$ so $\frac d{dq}\left(\prod_{j=1}^{D-2}\mathcal B_q\mathcal G_q\right)=0$; 
 where $\{l\} = \{l,\ldots,l\}$ and $\alpha_{j}= (D-j-1)/2$. By applying the known asymptotic behavior \cite{olver2010} of the gamma and digamma functions to $\Gamma(l+\frac{D}{2})$ and $\psi\left(l+\frac{D}{2}\right)$, respectively, it is straightforward to show that Eq. \eqref{eqI_cap1:SYgrado0} is consistent with Eq. \eqref{limit} in the limit $D \to \infty$. \\
 Then, for the spherical \textit{s}-states (i.e., $l=0$) we have the maximum value of the angular entropy
 \begin{equation}
 S \left[ {\cal{Y}}_{0,\left\lbrace 0 \right\rbrace},D\right]=\log \, \left[\frac{2\pi^{D/2}}{\Gamma(D/2)}\right],	
 \end{equation}
which provides the known values $\log(2\pi)$ and $\log(4\pi)$ for the two and three dimensional cases, respectively.
The explicit expression of the angular Shannon entropy for other specific electronic states can be obtained from Eqs. \eqref{shang} and \eqref{shang2} by use of modern symbolic algorithms of the type mentioned above, since these computer algebra systems are well equipped for the analytical manipulation and evaluation of generalized hypergeometric functions.\\

\section{ THE TOTAL SHANNON ENTROPY}
\label{sec:totsha}
Keeping in mind Eqs. \eqref{shaen1} and \eqref{denspos}, the total Shannon entropy of a $D$-dimensional hydrogenic state is given as
\begin{equation}
\label{shaen3}
S[\rho_{n,l,\{\mu\}},D]:=-\int_{\Delta} \rho_{n,l,\{\mu\}}(\vec{r})\log\rho_{n,l,\{\mu\}}(\vec{r})\, d\,\vec{r},\quad \Delta \subseteq \mathbb{R^{D}}.
\end{equation}
The numerical computation of this quantity is a formidable task, partially because of the number of integral singularities to avoid. Nevertheless, an effective method to perform it (which uses the coefficients of the recurrence relation of the involved orthogonal polynomials) has been given in the one-dimensional case \cite{buyarov2004}.\\
 For the analytical computation of $S[\rho_{n,l,\{\mu\}},D]$, which is the \emph{leitmotiv} of the present work, we have used a two-step method. First, following Eq. \eqref{shaen2} the total Shannon entropy is decomposed into the sum of the radial and angular Shannon entropies, and then we have encountered in the two previous sections compact expressions for these two quantities which can be symbolically solved by general-purpose computer program systems.\\
 In particular, one finds that the total Shannon entropy for the quasi-spherical states, $(n,n-1,\{n-1\}) = (n,\mu_1=\mu_2\ldots=\mu_{D-1}=n-1),$ of the $D$-dimensional hydrogenic system with nuclear charge $Z$  is given by

	\begin{align} \label{cir1}
		&S[\rho_{n,n-1,\{n-1\}},D] = D\,\log\left( \frac{e\,\sqrt{\pi}\,\eta}{2}\right)+\log\left(\frac{2\,\Gamma(2\,\eta+1)}{(n)_{\frac D2-1}}\right)+(n-1)\,\mathcal C_{D,n}-D\log Z,
	\end{align}
	where $\eta = n + \frac{D-3}{2}$ and $\mathcal C_{D,n}=\psi(\eta+\frac12)-\psi(n)-2\psi(2\eta+1)+2$. For $n=1$ one has the value

	\begin{align}\label{shannonRgs}\nonumber
		S[\rho_{1,0,\{0 \}},D]&=D\,\log \left( \frac{e\sqrt{\pi}}{4}\right)+\log\left(\frac{2\,(D-1)^D\,\Gamma(D)}{\Gamma\left(\frac D2\right)}\right)-D\log Z,
	\end{align}

for the total Shannon entropy for the ground state $(n,l,\{\mu\}) = (1,0,\{0\})$ of the $D$-dimensional hydrogenic system with nuclear charge $Z$. The second expression follows by taking into account that $\Gamma(\frac{D}{2})\Gamma(\frac{D+1}{2})=\sqrt{(2\pi)}2^{\frac{1}{2}-D}\,\Gamma(D).$  Moreover, from Eqs. \eqref{cir1}-\eqref{shannonRgs} we obtain the values

  	\begin{align}
 		S[\rho_{n,n-1},D=2] &= \log \left( \frac{\pi\,e^{2n}\,\Gamma(2n)\,(2n-1)^{2}}{8}  \right)-2(n-1)\psi(2n)-2\log Z, \nonumber\\
 		S[\rho_{n,n-1,n-1},D=3] &= \log\left[\pi\Gamma(n)^{2}n^{4}\right]+2n+\frac{1}{n}-(2n-2)\psi(n)-3\log Z\nonumber,
 \end{align}
 
 for the Shannon entropies of the quasi-circular states of the two and three-dimensional hydrogen ($Z=1$), respectively, and the known values
 \begin{align}\nonumber
 	S[\rho_{1,0},D=2] &= 2+\log\left(\frac{\pi}{8}\right)\\
 	S[\rho_{1,0,0},D=3] &= 3 + \log \pi
 \end{align}
for the Shannon entropies for the ground states of the two and three-dimensional hydrogen ($Z=1$), respectively. \\ 
Finally, note that in the limit $D \to \infty$ we obtain from Eq. \eqref{cir1} and the well-known asymptotical behavior \cite{olver2010} of the gamma function $\Gamma(x)$ and the digamma function $\psi(x)$, the values
\begin{align}
	& S[\rho_{n,n-1,\{n-1\}},D] \simeq D\log\left((D+2n-3)\,(D+2n-2)^\frac12\right) + D\log\left( \frac{\sqrt{\pi\,e}}{\sqrt 8\,Z}\right) -\log\left(\left(\frac{n}{e}\right)^{n-1}\sqrt 2\right),\quad D\to\infty
\end{align}
where the ($\log D$)-term exactly cancels. Thus we can write 
\begin{align}
	& S[\rho_{n,n-1,\{n-1\}},D] = \frac32\,D\,\log\,D + D\log\left( \frac{\sqrt{\pi\,e}}{\sqrt 8\,Z}\right) +\mathcal O(1).
\end{align}
%\textcolor{green}{
%\begin{align}
%	& S[\rho_{n,n-1,\{n-1\}},D] =\log\left(\frac{D^{2D}}{\Gamma\left(\frac D2\right)}\right)+D\log\left(\frac{\sqrt{\pi}}{4Z}\right)\ + \mathcal{O}(\log D)
%\end{align}
% I do not have completely clear how eq.(47) is derived, I obtain directly eq.(49).In any case, both are equivalent. }
%   \textcolor{green}{
% \begin{align}
%  	S[\rho_{n,n-1,\{n-1\}},D] = \log\left(\frac{D^{2D}}{\Gamma(\frac{D}{2})}\right) + D\log\left( \frac{\sqrt{\pi}}{4Z}\right)  + \mathcal{O}(D)\nonumber
%  \end{align}}
% and 
% \begin{equation}
% S[\rho_{1,0,\{0 \}},D]	 = \log\left(\frac{D^{2D}}{\Gamma(\frac{D}{2})}\right)+D\log\left( \frac{\sqrt{\pi}}{4Z}\right) + \mathcal{O}(D)
% \end{equation}
for the Shannon entropy of the high-dimensional hydrogenic quasi-spherical $(n,n-1,\{n-1\}) = (n,\mu_1=\mu_2\ldots=\mu_{D-1}=n-1)$ and ground ($n=1, l=0, \{\mu\}=\{0\}$) states, respectively, in a rigorous way. Most interesting it is to realize that these values are in agreement with (and improve) the recently obtained conjecture \cite{puertas2017b} that the total Shannon entropy $S[\rho_{n,l,\{\mu\}}]$ in position space for a general $(n,l,\{\mu\})$-state is given by
\begin{eqnarray}\label{eq:Cpshannon}
 S[\rho_{n,l,\{\mu\}},D]&=& \log\left(\frac{D^{2D}}{\Gamma\left(\frac D2\right)}\right)+D\log\left(\frac{\sqrt{\pi}}{4Z}\right)\ + \mathcal{O}(\log D)\nonumber\\
 &=& \frac32D\log D+D\log\left(\frac{\sqrt{e\pi}}{\sqrt 8Z}\right)+ \mathcal{O}(\log D),
\end{eqnarray}
which is a further checking of our results.

%%%%%%%%%%%%%%%%%%%%%%%%%%%%%%%%%%%%%%%%%%
\section{ CONCLUSIONS AND OPEN PROBLEMS}
\label{sec:conclu}
The analytical determination of the position-space uncertainty measures of Heisenberg-like (i.e., the radial expectation values of the position probability density $\rho(\vec{r})$) and entropic (Fisher information, R\'enyi entropy, Shannon entropy) types for all discrete stationary states of a $D$-dimensional hydrogenic system has been a fundamental issue since the early days of quantum mechanics. It requires to explicitly calculate these physical quantities in terms of $D$, the nuclear charge $Z$, and the state's hyperquantum numbers $(n,\mu_1,\mu_2,\ldots,\mu_{D-1})$. The Heisenberg-like measures were the first ones to be found by means of various methods \cite{ray1988,drake1990,andrae1997,tarasov2004,dehesa2010}, obtaining that they can be expressed \cite{zozor2011} by means of the univariate hypergeometric function $_3\mathcal F_2(x)$ evaluated at unity, i.e., $_3\mathcal F_2(1)$.\\
 The entropic uncertainty measures are much more complicated except the Fisher information, $F[\rho]=\int_{\mathbb{R}^{d}} \frac{|\nabla \rho|^{2}}{\rho}\,d\vec{r}$,  because of its close relationship with the kinetic energy operator \cite{romera2006}. Most recently, the R\'enyi entropy $R_{q}[\rho]$ (with integer $q$ greater than $1$), for all discrete multidimensional hydrogenic states \cite{toranzo2016c,puertas2018} has been found in a compact way by use of the Lauricella's multivariate hypergeometric function of type A evaluated at $1/q$, $F_{A}^{(2q)}(1/q,\ldots,1/q)$, and the Srivastava-Daoust' multivariate hypergeometric function evaluated at unity, $F_{1:1;\ldots;1}^{1:2;\ldots;2}(1,\ldots,1)$, respectively; the Lauricella function was coming from the radial part, and the Srivastava-Daoust function from the angular part of the quantity.\\
 In the present work we close this long-standing hydrogenic issue by expressing both radial and angular Shannon entropies of the multidimensional hydrogenic system in terms of the basic parameters {$(D, Z, n, \mu_1,\mu_2,\ldots,\mu_{D-1})$ of the system. This has been possible by use of the generalized hypergeometric funcions $F_{A}^{(s)}(1,\ldots,1)$ and $F_{1:1;\ldots;1}^{1:2;\ldots;2}(1,\ldots,1)$. The resulting compact expressions can be solved by general-purpose computer algebra systems such as Mapple and Mathematica and their variations \cite{koepf2014,gerdt2018} which can tackle the analytical problems involved in the evaluation of the aforementioned expressions for the Shannon entropy of the multidimensional hydrogenic systems considered in this work, including the manipulation of generalized hypergeometric functions. In particular, explicit expressions for the radial, angular and total Shannon entropies for the quasi-spherical $D$-dimensional hydrogenic states, which include the ground state, have been given. \\
 The extension of these position-space hydrogenic results to the momentum space has been done to a certain extent only. Indeed, the momentum Heisenberg-like measures (i.e., the radial expectation values of the momentum probability density $\gamma(\vec{p})$) \cite{hey1993,assche2000} and the momentum R\'enyi entropy $R_{q}[\gamma]$ (with integer $q$ greater than $1$) \cite{puertas2018} have been shown for all $D$-dimensional hydrogenic states to be expressed in terms of $D$, the nuclear charge $Z$, and the state's hyperquantum numbers by means of the univariate hypergeometric function $_5\mathcal F_4(1)$ and the Srivastava-Daoust' multivariate hypergeometric function $F_{1:1;\ldots;1}^{1:2;\ldots;2}(1,\ldots,1)$, respectively. However, the determination of the momentum Shannon entropy for all multidimensional hydrogenic states remain as an open problem.\\
 Finally, let us point out that the knowledge of the uncertainty measures for the extreme high-dimensional (pseudoclassical) and high-energy (Rydberg) hydrogenic cases is as follows. In the high-dimensional case the Heisenberg-like measures \cite{toranzo2016} and the entropic measures of R\'enyi type \cite{puertas2017b,sobrino2017} in both position and momentum spaces have been explicitly determined; but the position and momentum Shannon entropies of the high-dimensional hydrogenic states have not yet been rigorously calculated, although the dominant term has been conjectured \cite{puertas2017b}. In the high-energy case the Heisenberg-like measures \cite{aptekarev2010b,dehesa2010} and the Shannon entropy \cite{dehesa2010,lopez2013,toranzo2016b} in both position and momentum spaces as well as the position R\'enyi entropy \cite{toranzo2016b} have also been determined, but the momentum R\'enyi entropy of the Rydberg multidimensional hydrogenic states is not yet known.

\section*{Funding information}
{The work of J.S. Dehesa has been partially supported by the grant FIS2017-89349P of the Agencia Estatal de Investigaci\'on (Spain) and the European Regional Development Fund (FEDER). The work of N. Sobrino has been partially supported by the grant IT1249-19 of Basque Government and UPV/EHU.}

\section*{Author contributions}

J.S.D. and I.V.T. contributed in conceptualization, data curation, investigation, and writing of the original draft of the manuscript. D.P.C. and N. S. performed formal analysis, writing of the review, and edited the manuscript.

\section*{References}

%%%%%%%%%%%%%%%%%%%%%%%%%%%%%%%%%%%%%%%%%%%%%%%%%%%%%%%%%%%%%%%%%%%%%%%%%%%%%%%%%
% BIBLIOGRAPHY
%\bibliographystyle{ijqc}
%\bibliography{references}

%%%%%%%%%%%%%%%%%%%%%%%%%%%%%%%%%%%%%%%%%%
\end{document}